\documentclass[11pt]{article}
\usepackage{fullpage,amsthm}
\usepackage{amsmath,amssymb,amsfonts}
\usepackage{algorithmic}
\usepackage{graphicx}
\usepackage{textcomp}
\usepackage{xcolor}
\usepackage{longtable}
\usepackage{indentfirst}
\usepackage{subcaption}
\usepackage{hyperref}
\usepackage{makecell}
\usepackage{microtype}
\usepackage{lineno}
\usepackage{multirow,multicol}
\usepackage{cite}
\DeclareMathOperator*{\argmax}{arg\,max}
\DeclareMathOperator*{\argmin}{arg\,min}

\begin{document}
\baselineskip 12pt

\begin{center}
\textbf{\Large Carruthers Data Processing Pipeline: \\ Radiometric Responsivity Calibration} \\

\vspace{1.5cc}
{ \sc Alex Zhang$^{*1}$, Heather Filippini$^{2}$, Gonzalo Cucho-Padin$^{3}$, Lara Waldrop$^{1}$, John Clarke$^{4}$, Pratik Joshi$^{5}$, Parisa Karimi$^{6}$, Martin M. Sirk$^{7}$}\\

\vspace{0.3 cm}

{\small $^{1}$Department of Electrical and Computer Engineering, University of Illinois at Urbana-Champaign \\
$^{2}$Illinois Coordinated Sciences Laboratory, University of Illinois at Urbana-Champaign \\
$^{3}$ Goddard Space Flight Center, NASA \\
$^{4}$ Department of Astronomy and Center for Space Physics, Boston University \\
$^{5}$ Laboratory for Atmospheric and Space Physics, University of Colorado at Boulder \\
$^{6}$Formerly with Department of Electrical and Computer Engineering, University of Illinois at Urbana-Champaign \\
$^{7}$ Space Sciences Laboratory, University of California, Berkeley \\}
 \vspace{0.3 cm}
{\small $^{*}$Corresponding Author: alexmz2@illinois.edu}
 \end{center}

 \vspace{1.5cc}

\begin{abstract}
  \noindent The Carruthers Geocorona Observatory is NASA's first mission dedicated to investigating the fundamental nature of Earth’s exosphere and its dynamic response to space weather. Its primary payload, the GeoCoronal Imager, consists of two co-aligned broadband photometric imagers that support simultaneous, common-volume sensing of ultraviolet emission at 121.6 nm (Lyman-$\alpha$) by exospheric hydrogen atoms.  Accurate exospheric parameter retrieval from these images requires accurate knowledge of the instrument’s optical responsivity, which enables conversion of measured exospheric signal rates into the scientifically relevant quantity of emission radiance. The Carruthers mission achieves absolute sensitivity calibration through the acquisition of photometric measurements of stars and subsequent inversion of the observed fluxes to retrieve the wavelength-dependent responsivity across the passband for each imaging configuration. The retrieval algorithm performance is enhanced by its incorporation of an objective, algorithm-driven ranking criterion to systematically select target calibration stars from a stellar spectral library. The end-to-end workflow, from the stellar selection criterion to the passband inversion, is validated using synthetically generated stellar measurements.  These validation tests establish that our optical responsivity retrieval approach has high recovery fidelity, achieving error rates of $<5\%$ at the Lyman-$\alpha$ wavelength for all primary science imaging modes.
\vspace{0.95cc}

\parbox{24cc}{{\it Key words and Phrases}: Absolute Calibration, Carruthers Geocorona Observatory, GCI instrument, UV Stellar Dataset
}
\end{abstract}

\section{Introduction}

Optical remote sensing of the Earth's exosphere is a well established means of inferring exospheric state parameters, such as the density of its constituent hydrogen (H) atoms.   As the brightest optical emission line in the terrestrial airglow spectrum, the ultraviolet (UV) H emission line at 121.6 nm (Lyman-$\alpha$) has long been an attractive observation target, despite the need for space-based sensor deployment beyond 300 km altitude to avoid the nearly complete absorption of Lyman-$\alpha$ photons by molecular oxygen in Earth's upper atmosphere.   Relative to remote sensing from Earth-orbiting platforms using single-pixel photometers \cite{zoennchen2024} or scanning spectrographs \cite{bishop2001,joshi2018}, UV imaging systems provide superior constraints on the global exospheric emission scene.  The Carruthers Geocorona Observatory is NASA's first Heliophysics mission dedicated to routine exospheric imaging.   From its vantage in halo orbit around the Earth-Sun L1 Lagrange equilibrium point, which it reached shortly after launch in September, 2025, the Carruthers GeoCorona Imager (GCI) is acquiring high-resolution, high-cadence, wide-field images of Earth's Lyman-$\alpha$ emission.   These data support the Carruthers mission science objective to determine the nature and origin of global exospheric structure and dynamics, particularly in response to geomagnetic storms.  

The success of any radiometric investigation relies fundamentally on the accuracy of knowledge regarding the optical system’s response to the incident photons of interest. While the removal of instrument background signals and effects (such as dark current and flat field variations across the detector) is addressed in Zhang et al. \cite{Zhang26a}, and the subsequent isolation of the exospheric signal from background optical sources is detailed in Zhang et al. \cite{Zhang26c}, this paper describes the critical science data processing step of absolute sensitivity calibration of the isolated exospheric signal. Specifically, we present a novel, on-orbit approach to quantify the optical responsivity of the Carruthers GCI to enable conversion from the measured exospheric signal rate (in units of [Digital Numbers$ \; \text{s}^{-1}$], or [DN$ \; \text{s}^{-1}$]) to the scientifically-relevant Lyman-$\alpha$ radiance of the observed scene (in units of [photons$\; \text{s}^{-1} \; \text{cm}^{-2}$]).

Calibration of a sensor’s optical responsivity requires observations of a calibration source whose spectral radiance or incident photon flux is considered known, whether through modeling or measurement by other calibrated systems.  Ground-based laboratory calibration of optical sensors designed for space deployment is typically performed before launch, and the Carruthers GCI underwent extensive thermal/UV vacuum testing in late 2023 at the Centre Spatial de Liège (CSL) in Belgium \cite{Loicq16cslvaccuumfacility}. Most of the pre-launch integrated payload measurements, such as the camera point-spread function (PSF), optical component efficiencies, and end-to-end instrument responsivity, are derived from the CSL test campaign as described in detail by Rider et al. \cite{Rider24cslgcisetup}. Evaluation of the GCI's measured performance at CSL against its original design specifications is presented by Sirk et al.  \cite{carruthers_lab_cal_paper}. Throughout project inception and ground-testing, a project-level risk was maintained to avoid molecular contamination of the optics, as the GCI's UV passband of operation is highly contamination sensitive. More details on the contamination control plan can be found in Sirk et al. \cite{carruthers_lab_cal_paper}. On-orbit, the doors to the GCI imager were only opened well after the initial observatory outgassing period. Based on the gain at which the UV intensifiers are being operated and the expected stress on the instrument, it is anticipated that the instrument overall should not suffer more than 1\% degradation of throughput over the nominal 2-year prime mission.

However, as with most missions, on-orbit calibration is needed as a means to mitigate the risk of measurement bias introduced by potential post-launch changes in optical throughput due to thermal environment variability, deposition of contaminants, or degradation of optical components over time. Despite routinely appearing in the GCI scene as a very bright, extended source, the lunar disk is not an optimal ``standard candle" target owing to limited knowledge of its global surface albedo or the solar UV photon spectrum, its large variation in apparent brightness throughout the lunar cycle, and frequent lunar signal contamination by exospheric Lyman-$\alpha$ emission. The diffuse interplanetary hydrogen (IPH) background has been used for post-launch radiometric calibration of Lyman-$\alpha$ sensors onboard the MAVEN mission and the Hubble Space Telescope \cite{mayyasi2016}.  However, IPH-based calibration requires either independent knowledge of the incident solar photon flux at Lyman-$\alpha$ that excites the IPH emission, or coincident IPH observation by an independently calibrated sensor in support of cross-calibration.  For the Carruthers data processing pipeline, we instead utilize an alternative, stellar-based approach to radiometric calibration of the GCI that is motivated by the need to establish reliable, independent sensitivity calibration of Carruthers science data as early as possible to ensure accurate H density retrieval during the primary science phase of the mission. 

In space, a large number of stars emit brightly in the UV and are visible by the Carruthers GCI, and extensive archives of calibrated stellar measurements at UV wavelengths have been compiled over several decades by missions such as the International Ultraviolet Explorer (IUE) \cite{klinglesmith1979iue_image_processing_system}, the Hubble Space Telescope (HST) \cite{bohlin2019hubble_flux_cal}, the Ultraviolet and Infrared Atmospheric Spectrometer (SPICAM) \cite{bertaux2006spicam}, and the SOLar-STellar Irradiance Comparison Experiment (SOLSTICE) \cite{mcclintock2005solstice_mission}. Many stars in such archives are very hot, young and massive, with continuum black-body emission that peaks near the UV \cite{mondal2018uvit_demographic_of_stars}. Conventional stellar calibration, which uses observations of one UV star (or a few) to yield a scalar measurement of the instrument responsivity to the incident stellar photons, is most accurate for narrow-band optical systems, which are sensitive to a relatively small range of incident photon wavelengths over which incident stellar photon flux can be well approximated as constant \cite{holtzman1995performance_of_wfpc2_hubble, mondal2018uvit_demographic_of_stars, bohlin1979photometric_cal_iue_low_dispersion, snow2005solstice_two_solar_stellar_comparison, snow2013absolute_irradiance_moon_from_lasp}. However, broadband systems such as the GCI, which accrue detectable signal over a larger range of wavelengths, are more challenging to calibrate using these continuum stellar sources since both incident stellar photon flux and sensor responsivity vary significantly as a function of wavelength across the passband. Even when the incident stellar flux associated with a given scalar measurement of wavelength-integrated signal is well known, estimation of the underlying system responsivity as a function of wavelength is a severely under-determined problem. As a result, conventional calibration of broadband systems yields an effective (passband-averaged) system responsivity that can be difficult to interpret in terms of the desired responsivity at a particular wavelength of interest within the passband.  

This paper presents a general methodology for stellar calibration of broadband sensors that is capable of highly accurate estimation of UV sensor responsivity as a function of wavelength across the sensor passband. The method is robust even with only $30$ measurements of unique stellar targets, requiring far fewer observations than similar work by Weiler et al. \cite{weiler_passband_reconstruction}. Further, the passband considered in this paper focuses on the UV wavelengths, where stellar spectral data drawn from HST and IUE are far less reliable. The approach is based on the general signal processing principle of source separation \cite{rowe2002multivariate_source_separation}, whereby an ensemble of measurements of known stellar sources is used to constrain the linear system of equations governing the measurement model, which is then inverted using well-established regularization techniques.
We evaluate the algorithm's baseline performance using Monte Carlo simulations based on the numerical GCI image simulator \cite{Filippini26} to produce noisy, synthetic stellar data.  The results reported here assume all stars in the stellar spectral database are accessible at any time, bypassing the complex scheduling constraints typically involved in stellar target selection (such as target visibility, thermal/power constraints, and calibration time allocations). The impact of restricted stellar accessibility on target selection is a highly non-trivial optimization problem; a detailed analysis of these operational consequences is beyond the scope of this paper and is addressed comprehensively in Zhang et al. \cite{zhang2026deeprlfastlonghorizon}.  

\section{Payload Description}
\label{sec:payload_description}

This section provides a brief overview of the GCI optical design as it pertains to radiometric responsivity. A comprehensive description of the full payload architecture is available in Sirk et al. \cite{carruthers_lab_cal_paper} and a discussion of the detector electronics is provided in Zhang et al. \cite{Zhang26a}. The GCI consists of two co-aligned imagers (channels) which have differing fields of view. The Narrow Field Imager (NFI) images a $3.6^{\circ}$ Field-of-View (FOV) at a binned resolution of $1024 \times 1024$ pixels and is designed to provide high spatial resolution (better than 1 arcmin) near the Earth's limb where the exospheric radiance is expected to fall off rapidly. The Wide Field Imager (WFI) images an $18^{\circ}$ FOV at a binned resolution of $512 \times 512$ pixels and is designed to capture the entire exosphere with the sensitivity to detect the dimmest Ly-$\alpha$ signals at the outermost extent of the exosphere.

Despite their differing fields of view, the two channels share a similar, albeit physically independent, optical path for photon detection. Incident photons enter through an open entrance aperture, reflect off $\text{Al/MgF}_2$-coated mirrors (two for the NFI and four for the WFI) and pass through a 3 mm thick $\text{MgF}_2$ tube window featuring a 95\% transmissive Ni backside coating. The photons then strike a KBr photocathode, which liberates energetic photoelectrons with a 10\% quantum efficiency and is referred to as photoelectron ``events'' in the remainder of this paper. Accelerated by a $\sim 2100$ V two-plate chevron Microchannel Plate (MCP), these photoelectrons then initiate a cascading electron avalanche. The resulting electron cloud, comprising of $\sim 10^5$ electrons at the MCP exit, is accelerated across a 3000 V potential onto a phosphor screen, which subsequently emits green photons. Note that the number of electrons emitted for each incoming photon is modeled as a random variable following a non-standard empirically measured distribution, which we denote $G_{\text{mcp}}$. Finally, a fiber-optic taper demagnifies these photons onto a CMOS detector, generating a localized charge (referred to as ``electron-counts'', or [counts] in the remainder of this paper) that is read out by an Analog-to-Digital Converter (ADC) and converted into Digital Numbers (DN) with a constant gain factor $g_{\text{adc}}$. The full system acts as an intensified CMOS imaging system.

To achieve further spectral filtering, each channel employs a six-position filter wheel located upstream of the $\text{Al/MgF}_2$-coated mirrors. While the primary scientific objective is to isolate exospheric Ly-$\alpha$ emission at 1216{\AA}, fundamental FUV material limitations preclude the use of narrow-band filters capable of maintaining high Ly-$\alpha$ throughput while adequately suppressing the adjacent thermospheric Oxygen I emissions at 1304{\AA} and 1356{\AA}. To account for this unavoidable spectral blending, the filter wheels incorporate supplementary filters engineered to suppress Ly-$\alpha$ and independently measure these out-of-band terrestrial airglow signatures, whose combined radiances are estimated to be as bright as Ly-$\alpha$ radiance \cite{meier1991, mende2000fuvimagespacecraft}. The complete configuration and designated purpose of each filter are summarized in Table \ref{tab:filter_description}.

\begin{table}[htbp]
    \begin{center}
    \caption{Description of the $6$ filters in each filter wheel equipped by the two channels.}
    \label{tab:filter_description}
        \begin{tabular}{ |c|c|c| }
            \hline
            Filter & Description & Purpose \\
            \hline
            \hline
            Blocked & Aluminum blocking disk & \makecell{Zero transmissivity at all wavelengths. \\ Used for calibration purposes only.} \\
            \hline
            Open & No filter & Transmissivity of unity at all wavelengths. \\
            \hline
            LyaN & \makecell{Ly-$\alpha$ narrow band filter \\ (Acton F122-N, \\ peak transmissivity 22\%)} & \makecell{Nominal science filter, \\ supports narrowband Ly-$\alpha$ transmission.} \\
            \hline
            LyaX & \makecell{Ly-$\alpha$ extra narrow band filter \\(Acton F122-XN, \\ peak transmissivity 11\%)} & \makecell{Alternative science filter if exosphere \\ is brighter than model predictions. \\ Supports narrowband Ly-$\alpha$ transmission.} \\
            \hline
            CaF2 & 3mm thick calcium fluoride window & \makecell{Suppresses Ly-$\alpha$ transmission \\ (and only transmits longer wavelengths) \\ for out-of-band measurement and subtraction.} \\
            \hline
            SrF2 & 3mm thick strontium fluoride window & \makecell{Suppresses Ly-$\alpha$ transmission \\ (and only transmits longer wavelengths) \\ for out-of-band measurement and subtraction.} \\
            \hline
        \end{tabular}
    \end{center}
\end{table}

\section{Responsivity Definition}
\label{sec:resp_definition}

\begin{table}[htbp]
    \centering
    \caption{Nomenclature of symbols and parameters}
    \label{tab:responsivity_symbols}
    \renewcommand{\arraystretch}{1.2}
    \begin{tabular}{l p{0.25\textwidth} p{0.5\textwidth}}
        \hline
        \textbf{Symbol} & \textbf{Units} & \textbf{Description} \\
        \hline
        $a_c$ & [cm$^2$] & Channel $c$'s aperture area \\
        $b$ & N/A & Filter index \\
        $\mathbb{E}[\cdot]$ & N/A & Expected value operator \\
        $f(\lambda, i, j, t)$ & [photons$\; \text{s}^{-1}\; \text{cm}^{-2}\; \text{\AA}^{-1}$] & Stellar photon flux \\
        $g_{\text{adc}}$ & [DN$ \; \text{count}^{-1}$] & ADC gain constant \\
        $G_{\text{mcp}}$ & \raggedright [count$\; \text{event}^{-1}$] & Microchannel plate (MCP) gain distribution (measured empirically) \\
        $h(i,j)$ & N/A & Instrument Point Spread Function (PSF) \\
        $i, j$ & N/A & Pixel indices in the Field-of-View (FOV) \\
        $i_m, j_m$ & N/A & Pixel coordinate onto which star $m$ is projected \\
        $k$ & N/A & Image frame index \\
        $\ell(\lambda, i, j, t)$ & [photons$\; \text{s}^{-1}\; \text{cm}^{-2}\; \text{\AA}^{-1}$] & Total spectral radiance of the diffuse scene \\
        $m$ & N/A & Index of the calibration star in the target database \\
        $n_{\text{frame}}$ & N/A & Total number of image frames \\
        $r_b(\lambda)$ & [DN$\; \text{photon}^{-1} \; \text{cm}^2$] & Spectral responsivity of filter $b$ as a function of wavelength $\lambda$ \\
        $S_{b}(i, j)$ & [DN] & Random variable representing the signal value of pixel $i,j$ in the FOV in an image taken with filter $b$ with all instrument backgrounds perfectly removed \\
        $S_{m, b}$ & [DN] & Random variable representing the total integrated signal of star $m$ in filter $b$ \\
        $t_{\text{frame}}$ & [s] & Integration time duration of a single frame \\
        $t_{\text{int}}$ & [s] & Total integration time \\
        $t_k$ & [s] & Start time of frame $k$ \\
        $\delta(\cdot)$ & N/A & Dirac delta function \\
        $\varepsilon(\lambda)$ & [events$\; \text{photon}^{-1}$] & Wavelength-dependent open system efficiency \\
        $\lambda$ & [\AA] & Wavelength \\
        $\Omega_c$ & [sr] & Channel $c$'s pixel solid angle \\
        $\tau_b(\lambda)$ & N/A & Wavelength-dependent filter transmissivity for filter $b$ \\
        \hline
    \end{tabular}
\end{table}

Table \ref{tab:responsivity_symbols} lists the symbols and definitions used in the remainder of this paper. Throughout this paper, upper-case letters (English or Greek) denote random variables, upper-case letters with an arrow, such as $\vec{X}$, denote random vectors, bold upper-case letters (English or Greek), such as $\boldsymbol{X}$ denote matrices or sets, lower-case letters with an arrow, such as $\vec{x}$, denote vectors, and lower-case letters denote constants or indices. Variables $\epsilon$ that are a function of wavelength are denoted $\epsilon(\lambda)$, while variables $\epsilon$ that are a function of pixel position are denoted $\epsilon(i, j)$, where $i$ denotes the row index ($y$ direction) and $j$ denotes the column index ($x$ direction).

Optical responsivity of filter $b$ as a function of wavelength, $r_b(\lambda)$, can be defined by starting with the expected value of the signal $S_b$ in pixel $i,j$, denoted $\mathbb{E}\left[ S_{b}(i,j) \right]$.  When observing a diffuse scene, characterized by spectral radiance $\ell(\lambda, i, j, t)$ in units of [photons$ \; \text{s}^{-1} \; \text{cm}^{-2} \; \text{\AA}^{-1}$] (assumed to be isotropically radiated over $4\pi$ steradians), the expected signal measured in a pixel within the channel field-of-view (FOV) is given by: 

\begin{equation}
    \label{eq:fov_full_eq}
    \mathbb{E}\left[ S_{b, \ell}(i,j) \right] = \sum_{k = 1}^{n_{\text{frame}}} g_{\text{adc}}\mathbb{E}[G_{\text{mcp}}] \frac{a_c\Omega_c}{4\pi} \int_{t_k}^{t_k + t_{\text{frame}}} \int_{0}^{\infty} \varepsilon(\lambda)\tau_b(\lambda)(\ell(\lambda, i, j, t) * h(i,j)) d\lambda dt
\end{equation}

Here, $G_{\text{mcp}}$ is a random variable representing the number of electrons emitted from the MCP due to a single detected UV photon; thus, the expected signal value in an image pixel depends only on the expected value of the random variable representing MCP gain, denoted $\mathbb{E}[G_{\text{mcp}}]$ in Equation \ref{eq:fov_full_eq}. In contrast, $g_{\text{adc}}$ is a constant gain multiplier applied by the Analog-to-Digital Converter (ADC) and therefore does not require the expected value notation. The geometric scaling factor $\frac{a_c\Omega_c}{4\pi}$ accounts for the geometric collection efficiency of channel $c$: it combines the channel's aperture area $a_c$ and the channel pixel's solid angle fraction $\frac{\Omega_c}{4\pi}$. The convolution with the instrument point-spread-function (PSF) $h(i, j)$ represents the spatial blurring of the incoming light caused by the instrument's optics and spacecraft jitter, with $h(i,j)$ found to be well-modeled by a Voigt profile and theoretically widened by the expected jitter, thermal distortions, and other pointing errors \cite{carruthers_lab_cal_paper}.  In the above expression, and for the validation testing reported here, we assume that all instrument backgrounds are removed with perfect accuracy; the reader is referred to Zhang et al. \cite{Zhang26c} for a detailed discussion on instrument background removal accuracy. 

In contrast, observing a stellar point source characterized by stellar photon flux $f(\lambda, i, j, t)$ in units of [photons$ \; \text{s}^{-1} \; \text{cm}^{-2} \; \text{\AA}^{-1}$] requires slightly different treatment. The expected signal measured in a pixel within the channel FOV from a stellar point source is given by:
\begin{equation}
    \label{eq:fov_star_full_eq}
    \mathbb{E}\left[S_{b, f}(i,j) \right] = \sum_{k = 1}^{n_{\text{frame}}} g_{\text{adc}}\mathbb{E}[G_{\text{mcp}}] a_c \int_{t_k}^{t_k + t_{\text{frame}}} \int_{0}^{\infty} \varepsilon(\lambda)\tau_b(\lambda)(f(\lambda, i, j, t) * h(i,j)) d\lambda dt
\end{equation}
The only differences between Equations \ref{eq:fov_full_eq} and \ref{eq:fov_star_full_eq} are the exchange of $\ell(\lambda, i, j, t)$ for $f(\lambda, i, j, t)$ and the removal of the fractional solid angle of a channel pixel $\frac{\Omega_c}{4\pi}$. The fractional solid angle, along with the aperture area, is needed to relate measured signals to their diffuse emission sources, and the science data pipeline accounts for this channel pixel geometric factor in the absolute calibration of the diffuse exospheric and IPH Lyman-$\alpha$ scenes. However, the signal measured when observing a stellar point source is dependent only on a telescope's aperture area and not on the camera field-of-view or detector resolution, both of which determine $\Omega_c$. Thus, the solid angle fraction $\frac{\Omega_c}{4\pi}$ must be omitted from Equation \ref{eq:fov_star_full_eq}.

The formulation of optical responsivity selected for this paper represents the quantity directly retrievable by absolute calibration using stellar sources. Therefore, the GCI's optical responsivity, $r_b(\lambda)$, relates the signal rate in [DN s$^{-1}$] to the stellar photon flux of the scene, $f(\lambda, i, j, t)$, and thus has units of [DN$ \; \text{photon}^{-1} \; \text{cm}^2$]. The responsivity across the passband is strongly wavelength-dependent and is defined by aggregating the relevant factors in equation ~\ref{eq:fov_star_full_eq}:
\begin{equation}
    \label{eq:resp_funct_def}
    r_{b}(\lambda) = a_c g_{\text{adc}}\mathbb{E}[G_{\text{mcp}}] \varepsilon(\lambda)\tau_b(\lambda)
\end{equation}

In order to utilize the optical responsivity $r_b(\lambda)$ derived from stellar point sources when relating measured signals to their diffuse emission sources, the fractional solid angle $\frac{\Omega_c}{4\pi}$ must not be forgotten. To be explicit, Equations \ref{eq:fov_full_eq} and \ref{eq:fov_star_full_eq} are rewritten below in terms of $r_b(\lambda)$:

\begin{equation}
    \label{eq:fov_full_eq_with_rep}
    \mathbb{E}\left[ S_{b, \ell}(i,j) \right] = \sum_{k = 1}^{n_{\text{frame}}} \frac{\Omega_c}{4\pi} \int_{t_k}^{t_k + t_{\text{frame}}} \int_{0}^{\infty} r_b(\lambda)(\ell(\lambda, i, j, t) * h(i,j)) d\lambda dt
\end{equation}
\begin{equation}
    \label{eq:fov_star_full_eq_with_rep}
    \mathbb{E}\left[ S_{b, f}(i,j) \right] = \sum_{k = 1}^{n_{\text{frame}}} \int_{t_k}^{t_k + t_{\text{frame}}} \int_{0}^{\infty} r_b(\lambda)(f(\lambda, i, j, t) * h(i,j)) d\lambda dt
\end{equation}

\begin{figure}[ht]
    \begin{subfigure}{0.48\textwidth}
        \includegraphics[width=\textwidth]{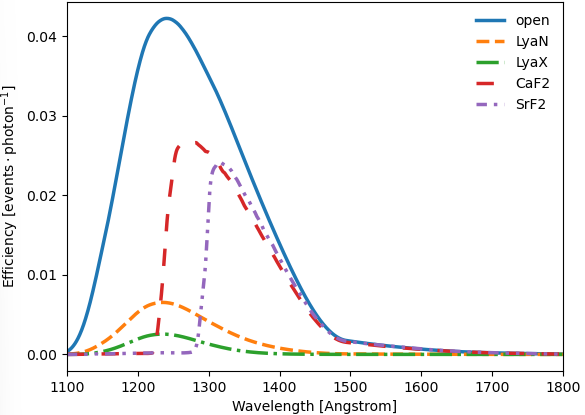}
        \caption{WFI}
    \end{subfigure}
    \hfill
    \begin{subfigure}{0.48\textwidth}
        \includegraphics[width=\textwidth]{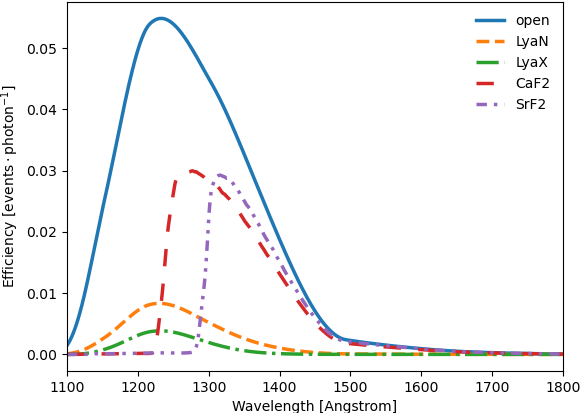}
        \caption{NFI}
    \end{subfigure}
    \caption{Nominal system efficiency curves for the WFI (a) and NFI (b) channels across all integrated filters, derived from pre-launch ground testing. System efficiency $\varepsilon(\lambda)\tau_b(\lambda)$ represents the dimensionless probability of an incident photon triggering a detected event, bounded between 0 and 1, and is related to the instrument's absolute responsivity $r_b(\lambda)$ by known constant scaling factors. Table \ref{tab:filter_description} provides a detailed description of the five filters listed in the legend. The sixth filter, the blocked filter, is omitted from this plot as the blocked system efficiency is zero at all wavelengths. \label{fig:lab_responsivity}}
\end{figure}

Throughout this paper, while our overarching objective is to characterize the absolute responsivity $r_b(\lambda)$, we present our results in terms of filter $b$'s system efficiency $\varepsilon(\lambda)\tau_b(\lambda)$ [event $\; \text{photon}^{-1}$], which is bounded between 0 and 1 and represents the end-to-end probability of a photon incident on the aperture producing a recorded event. Because system efficiency and absolute responsivity are related by wavelength independent scaling factors, operating in the efficiency domain provides a more intuitive framework for validating our retrieval algorithms without sacrificing physical meaning. Figure \ref{fig:lab_responsivity} depicts these nominal system efficiency curves for all filters across both channels derived from laboratory measurements and represents the end-to-end performance of the fully integrated instrument measured during pre-launch ground testing \cite{carruthers_lab_cal_paper}.

\section{Responsivity Retrieval}
\label{sec:retrieval_algo}

This section discusses the responsivity retrieval algorithm. We first eliminate the time-dependence of the incident stellar flux in Equation \ref{eq:fov_star_full_eq_with_rep}, as we assume that all stars are photometrically stable over their measurement period. With all time-dependence removed, the integration over time and the summation over image frames can both be replaced by a multiplication with integration time $t_{\text{int}}$, in units of [s]:
\begin{equation}
    \mathbb{E}\left[S_{b, f}(i,j)\right] = t_{\text{int}} \int_{0}^{\infty} r_b(\lambda)(f(\lambda, i, j) * h(i,j)) d\lambda
\end{equation}
In this context, no other sources of photons are assumed present, as stellar measurements are not taken when stellar signals are obscured by Earth's exosphere, the Moon, or any of the outer planets, and any other sources of photons are assumed to have been removed without error. The reader is referred to Zhang et al. \cite{Zhang26a} for a detailed discussion on photon background removal accuracy. 

For simplicity, we assume that there is only a single star in the image FOV whose index in the calibration star target database is given by $m$. This assumption allows the ensemble scene $f(\lambda, i, j)$ to be simplified as $f_m(\lambda) \delta(i - i_m)\delta(j - j_m)$, where pixel $i_m, j_m$ is the pixel onto which star $m$ is projected in the camera frame.
\begin{equation}
    \mathbb{E}\left[S_{b, m}(i,j) \right] = t_{\text{int}} \int_{0}^{\infty} r_b(\lambda)\left(f_m(\lambda) \delta(i - i_m)\delta(j - j_m) * h(i,j)\right) d\lambda
\end{equation}
The convolution can now be simplified as follows:
\begin{equation}
    \mathbb{E}\left[S_{b, m}(i,j)\right] = t_{\text{int}}\int_{0}^{\infty} r_b(\lambda)f_m(\lambda)h(i - i_m,j - j_m) d\lambda
\end{equation}
Finally, the signal is integrated over the full spatial domain $(i, j)$ to recover the total source flux. By exploiting the PSF normalization condition ($\sum_{ij} h(i, j) = 1$), this spatial summation accounts for all distributed light, rendering the theoretical flux measurement independent of the star's specific centroid location. In practice, however, the summation is restricted to a finite aperture that encompasses the non-negligible wings of the PSF to minimize noise. Furthermore, stellar measurements are confined to a consistent location on the detector to mitigate the effects of spatial non-uniformity in the detector response. The resulting expected value for the total signal $\mathbb{E}\left[ S_{b, m} \right]$ of star $m$ in filter $b$, in units of [DN], is
\begin{equation}
    \mathbb{E}\left[S_{b, m}\right] = t_{\text{int}} \int_{0}^{\infty} r_b(\lambda)f_m(\lambda)d\lambda
\end{equation}

While the preceding expression models a single stellar source, numerical evaluation requires truncation and discretization. The integration limits are restricted to the 1100-1800{\AA} range, which corresponds to the instrument's effective passband. Because the stellar spectral radiances utilized are sourced from the IUE SWP catalog, the available data truncates at roughly 1950{\AA}. Analysis of the 1800-1950{\AA} spectral tail for the cooler stars in our database confirms that its contribution to the total expected signal is negligible due to the rapid decline in the KBr photocathode's quantum efficiency. Therefore, restricting the numerical integration to 1800{\AA} captures the entirety of the meaningful signal.

Within the 1100-1800{\AA} window, the integral is approximated as a discrete sum with a 0.25{\AA} spectral sampling interval, which was chosen as a resolution sufficient to resolve the variations in responsivity (see Figure \ref{fig:lab_responsivity}) without introducing aliasing errors. The sampling density chosen also exceeds the native spectral resolution of all candidate calibration stars in the available databases. Going forward, we distinguish continuous functions (e.g., $r_b(\lambda)$) from discretized vectors (e.g., $\vec{r}_b$), strictly utilizing vector notation post-discretization.
\begin{equation}
    \mathbb{E}\left[S_{b, m}\right] = t_{\text{int}} \vec{f}_m^T \vec{r}_b
\end{equation}

Finally, the individual expressions for the full set of $n$ stars are concatenated into a single matrix formulation:
\begin{equation}
    \label{eq:final_matrix_eq}
    \begin{bmatrix}
        \mathbb{E}\left[S_{b, \text{star} \ 1}\right] / t_{\text{int}, 1} \\
        \mathbb{E}\left[S_{b, \text{star} \ 2}\right] / t_{\text{int}, 2} \\
        \vdots \\
        \mathbb{E}\left[S_{b, \text{star} \ n}\right] / t_{\text{int}, n} \\
    \end{bmatrix} = \begin{bmatrix}
    \cdots & \vec{f}_{\text{star} \ 1} & \cdots \\
    \cdots & \vec{f}_{\text{star} \ 2} & \cdots \\
    \cdots & \vdots & \cdots \\
    \cdots & \vec{f}_{\text{star} \ n} & \cdots \\
    \end{bmatrix} \begin{bmatrix}
        \vdots \\ \vec{r}_b\\ \vdots
    \end{bmatrix}
\end{equation}
The above can be rewritten in matrix form to obtain $\vec{S}_{\text{stars}} = \boldsymbol{F}\vec{r}$. Note that in operational practice, the integration time $t_{\text{int}, m}$ varies with star $m$ and is tailored (using the measured lab responsivities reported in terms of system efficiency in Figure \ref{fig:lab_responsivity}) to ensure that a minimum target Signal-to-Noise Ratio (SNR) of 40 is achieved across all measurements. Because dimmer sources are observed for longer durations to match the data quality of brighter sources, and the matrix formulation (Equation \ref{eq:final_matrix_eq}) normalizes the signal by $t_{\text{int}}$, we naturally place all calibration targets on an equal statistical footing without requiring magnitude-based weighting.  Our   choice for target SNR is based on balancing nominal operational staring times, derived from simulated stellar data, with operational requirements on Earth nadir-staring efficiency, the availability of suitably bright candidate stars, and extensive responsivity retrieval performance testing under these constraints. 

On-orbit, we do not have access to the true expected value on the left-hand side of Equation \ref{eq:final_matrix_eq}; thus, the inverse problem is to recover responsivity $\vec{r}$ given a noisy instance of the random vector $\vec{S}_{\text{stars}}$. However, the linear system is severely underdetermined: the number of stars observed (30) corresponds to the number of rows in $\boldsymbol{F}$, while the number of samples in the wavelength-domain ($\sim 2800$) corresponds to the number of columns in $\boldsymbol{F}$. A prior $\vec{p}$ is therefore introduced to constrain the final solution. The prior is selected to be the measured lab responsivities as displayed in Figure \ref{fig:lab_responsivity}.

The analysis below is divided into two distinct categories based on filter type: bandpass filters (Open, LyaN, and LyaX) and longpass filters (CaF2 and SrF2). This separation is necessary for two primary reasons:
\begin{enumerate}
    \item Geometric Consistency: Filters within the same class share similar shapes. Because the bandpass filters follow a consistent profile, regularization techniques applied to one are generally effective for the others. The same logic applies to the longpass group.
    \item Thermal Responsivity: The two classes behave differently under temperature fluctuations. The longpass filters exhibit a ``cut-on'' wavelength, defined as the point where transmission rises sharply from zero (approximately 1220{\AA} for CaF2 and 1300{\AA} for SrF2). This cut-on wavelength is known to shift as a function of temperature \cite{laufer_filter_cutoff}. Consequently, data retrieval models for these filters must account for thermal drift. On the other hand, the bandpass filters are thermally stable; laboratory measurements found that no shifts with temperature were detected within the measurement uncertainty.
\end{enumerate}

The remainder of this section is divided as follows: Subsection \ref{sec:target_selection} discusses stellar target selection. Subsection \ref{sec:validation_framework} describes the framework used to validate these algorithms. Finally, subsection \ref{sec:results_bandpass} discusses the retrieval algorithm performance results for the bandpass filters, while subsection \ref{sec:results_longpass} discusses the retrieval algorithm performance results for the longpass filters.

\subsection{Stellar Target Selection}
\label{sec:target_selection}

Stellar target selection is required to test the responsivity retrieval algorithm. However, a proper treatment of stellar target selection involves many scheduling constraints, which are out of scope for this paper (see Zhang et al.  \cite{zhang2026deeprlfastlonghorizon} for a full treatment). In order to obtain a baseline for the performance of the algorithms presented in this section, the validation test assumes that any star in the stellar spectra database is observationally feasible. However, the number of stars used to constrain the responsivity retrieval in the algorithm performance evaluation presented here is limited to 30.  This number of distinct stellar targets aligns with the mission’s minimum operational requirement that was established in the design phase of the mission based on comprehensive simulation and validation testing, which is omitted here for brevity.  

The matrix $\boldsymbol{F}$ is constructed by populating its rows with the stellar flux spectra of the $30$ calibration targets. In order to facilitate the most accurate retrieval of $\vec{r}_b$, the condition number of $\boldsymbol{F}$ should be as small as possible \cite{trefethen2022numerical}. Defined mathematically as the ratio of the matrix's maximum singular value to its minimum singular value, the condition number dictates the numerical stability of the inversion process, bounding the worst-case measurement noise and numerical error amplification after inversion. Due to the number of stars involved, it is computationally infeasible to evaluate the condition number of every possible subset of $30$ stars out of the $858$ available in the candidate  database to find the subset that minimizes the condition number. Instead, the coherence of the rows of the matrix is chosen as the metric to be minimized when selecting stellar targets. Coherence is defined as

\begin{equation}
    \label{eq:coherence_formula}
    \mu(\boldsymbol{F}) = \max_{u\neq v} \frac{\|\langle \boldsymbol{F}_u, \boldsymbol{F}_v\rangle\|}{\|\boldsymbol{F}_u\|\|\boldsymbol{F}_v\|}
\end{equation}
where $\boldsymbol{F}_u$ denotes the $u$-th row of matrix $\boldsymbol{F}$. The coherence metric defined above quantifies the similarity between rows; the numerator is minimized as the row vectors approach orthogonality. Notably, the inner products and induced norms in Equation \ref{eq:coherence_formula} are not restricted to the standard Euclidean ($L_2$) definition. Given the availability of a high-fidelity prior, $\vec{p}$, namely, the lab-based measurement of the system responsivity acquired before launch, a weighted inner product space with $\vec{p}$ as the weighting vector can instead be adopted. The weighted inner product thus ensures prioritization of row orthogonality only at wavelengths where the system responsivity is known to be significant. By discounting contributions from wavelengths where the responsivity is negligible, the conditioning of the inversion is improved specifically where the signal is known to exist.

Coherence can be minimized greedily via iteratively selecting stars and adding them to the matrix $\boldsymbol{F}$, choosing the star that minimizes coherence at each decision step. However, picking the first star presents a problem as there are no other rows to compare against. The greedy algorithm arbitrarily chooses to find the first star $m$ by solving the following maximization problem
\begin{equation}
    \argmax_{m} \frac{\|\langle \vec{f}_m, \vec{p}\rangle\|}{\| \vec{f}_m \|\;\|\vec{p}\|}
\end{equation}

In order to simulate measurements of stars, the mean and variance of the stellar measurement are directly calculated using the numerical formulation derived in Filippini et al. \cite{Filippini26}. However, the formulae are modified to assume perfect background subtraction. In particular:
\begin{itemize}
    \item All instrument backgrounds and other photon backgrounds besides the target star are removed when calculating the mean of the stellar measurement.
    \item The variance of the stellar measurement is calculated with all instrument background terms and only the InterPlanetary Hydrogen (IPH) and star included as photon sources (other photon sources will not be in the background of stellar measurements during on-orbit calibration operations). The IPH photon source is simulated for a PSF-sized kernel, then integrated to obtain the IPH photon background that will be present in the stellar measurements. This integration is necessary to account for the cumulative IPH signal distributed across the full spatial extent of the PSF that will be integrated when obtaining the stellar measurement on-orbit.
\end{itemize}

For the algorithm performance validation described in Section \ref{sec:validation_framework}, using the stellar fluxes in the calibration database to create synthetic stellar measurements would assume perfect knowledge of stellar fluxes, which is not necessarily true. Thus, a probabilistic model of stellar fluxes is constructed to maintain a conservative estimate of performance. Let $\vec{F}_m$ denote the multivariate random variable for stellar flux $m$. The model for $\vec{F}_m$ is chosen to be a multivariate Gaussian with mean equal to the stellar flux $\vec{f}_m$ in the dataset. In the absence of knowledge of higher-order statistical moments or sufficient data to empirically derive a complex prior, the Gaussian distribution represents the maximum entropy assumption \cite{jaynes1957information}, which minimizes the introduction of unwarranted structural bias into the model while rigorously accounting for the known first and second moments. The model assumes two sources of covariance:
\begin{itemize}
    \item Random error, which is a diagonal covariance matrix $\boldsymbol{\Sigma}_r$ whose diagonal is given in the stellar dataset.
    \item Systematic error, which is a rank-1 covariance matrix $\boldsymbol{\Sigma}_s$, created as the outer-product of 1\% of the stellar flux $\vec{f}_m$ with itself.
\end{itemize}
The two sources of error are assumed to be independent. Thus, the stellar flux to be used when calculating the mean and variance of the synthetic stellar measurement is given by a random draw from this multivariate Gaussian random variable $\vec{F}_m$ with mean $\vec{f}_m$ and covariance $\boldsymbol{\Sigma}_r + \boldsymbol{\Sigma}_s$. Note that the stellar flux matrix $\boldsymbol{F}$ is still constructed using the mean flux $\vec{f}_m$ found in the stellar database.

\subsection{Validation Framework}
\label{sec:validation_framework}

Validation of the stellar calibration algorithm is assessed independently for each channel/filter combination. For each of these combinations, the $30$ stars to be observed are found using the algorithm described earlier in the previous section (i.e. by minimizing coherence using a weighted dot product with weight equal to the prior). Next, the stellar flux matrix $\boldsymbol{F}$ is constructed using the selected stellar fluxes. To establish a uniform computational domain, all stellar fluxes and the prior are interpolated onto a common spectral grid spanning 1100--1800{\AA} with a step size of 0.25{\AA}. Justification for the spectral bounds and the sampling density was discussed above in Section \ref{sec:retrieval_algo}.

Next, we construct the ground-truth responsivity. To avoid false-positive results in our retrieval validation tests, the ground-truth profiles are generated by applying synthetic perturbations to the pre-launch laboratory measurements rather than using the pre-launch laboratory measurements directly. Introducing these variations demonstrates the algorithm's robustness to potential on-orbit behavioral shifts. For the open, LyaN, and LyaX filters, the laboratory measurements were uniformly scaled by a factor of 0.9. Note that this factor represents the cumulative degradation over a full one-year period between annual stellar calibration epochs; following each annual observation, the newly measured degraded responsivity is utilized as the prior for the subsequent year. Long-term orbital data from spaceborne ultraviolet instruments demonstrate that nominal optical throughput degradation at UV wavelengths is typically on the order of 1\% per year, as documented by calibration studies for the Swift Ultraviolet/Optical Telescope \cite{Breeveld_2011}. Therefore, our adopted 10\% annual loss is not intended as an expected physical performance metric. Rather, it serves as a conservative upper bound on potential throughput loss within a year, ensuring the responsivity retrieval algorithm can yield a reliable solution even in the presence of substantial photometric degradation.

For the CaF2 and SrF2 filters, the laboratory measurements were both uniformly scaled by 0.9 and shifted either 0{\AA}, 5{\AA} left, or 5{\AA} right. These shifts are introduced because the crystalline fluoride substrates (CaF2 and SrF2) are known to show filter edge drifts as a function of temperature \cite{laufer_filter_cutoff}. The thermal drift of the filter edge was measured during pre-launch laboratory calibrations by performing filter transmission measurements for the CaF2 and SrF2 filters with a fine sampling grid of 2{\AA} at the transmission cut-off edge while the temperature of the instrument was slowly varied from 35C to 5C. The measured thermal drift of the filter edges for the CaF2 and SrF2 filters was found to be $+0.451 \pm 0.032$ and $+0.477 \pm 0.056$ [{\AA}$\; \text{K}^{-1}$], respectively \cite{carruthers_lab_cal_paper}.

To validate the responsivity retrieval algorithms detailed in the following two sections, we aggregated and analyzed the results of $50$ independent trials for each channel/filter combination. Each trial utilized a unique set of noisy synthetic measurements for the $30$ selected stars, simulated according to the noise model outlined in Section \ref{sec:validation_framework}. The aggregate percentage error statistics presented below compare the retrieved responsivities across all $50$ trials against the ground truth. Specifically, the percent error is calculated as
\begin{equation}
    \label{eq:perc_err_def}
    \vec{E} = 100 \frac{\hat{\vec{r}} - \vec{r}}{\vec{r}}
\end{equation}
where $\vec{r}$ is the ground-truth responsivity, $\hat{\vec{r}}$ is the retrieved responsivity for a single trial, and the division is understood to be elementwise. The percent error is denoted by a capital letter to indicate it is a random vector, whose randomness originates from $\hat{\vec{r}}$, which varies across independent trials depending on the noise model. The empirical mean and standard deviations over $50$ trials at a particular wavelength $\lambda$ are denoted $\hat{\mathbb{E}}[\vec{E}_{\lambda}]$ and $\hat{\sigma}[\vec{E}_{\lambda}]$ respectively. This analysis focuses specifically on the retrieved responsivity at 1216{\AA} for the open, LyaN, and LyaX filters across both channels, as well as at 1304{\AA} and 1356{\AA} for the NFI open, NFI LyaN, NFI CaF2, and NFI SrF2 filters. These specific configurations represent critical channel, filter, and wavelength combinations necessary to achieve the mission's scientific objectives \cite{Waldrop26}. Although the algorithm yields responsivities at supplementary wavelengths as a secondary output, these values are not subject to formal retrieval accuracy requirements nor used in the Carruthers data processing pipeline. Consequently, the subsequent validation for these additional wavelengths remains strictly qualitative.

\subsection{Algorithms and Results: Bandpass Filters}
\label{sec:results_bandpass}

Recall from Section \ref{sec:retrieval_algo} that $\vec{S} = \boldsymbol{F}\vec{r}$, but access is limited to a noisy instance of $\vec{S}$ and $\boldsymbol{F}$ is a wide matrix. Thus, the linear inverse problem is ill-posed and direct inversion of $\boldsymbol{F}$ is impossible. A generalized Tikhonov regularization framework is consequently adopted (cf. Aster et al. \cite{aster2018parameter}).  The solution space is constrained using prior, pre-launch, lab-based knowledge of the responsivity profile, denoted $\vec{p}$.

For the bandpass filters, we formulate an objective function that balances the data misfit with other loss terms. The expression to be minimized is given as
\begin{equation}
\label{eq:bandpass_minimizer}
    \hat{\vec{r}} = \argmin_{\vec{r}} |\vec{S} - \boldsymbol{F}\vec{r}|_2^2 + \gamma_1 |\vec{r}|_{\vec{p}}^2 + \gamma_2 |\vec{p} \circ \boldsymbol{D}(\vec{r} - \vec{p})|_2^2
\end{equation}

Here, $|\vec{r}|_{\vec{p}}^2 = \sum_u \frac{r_u^2}{p_u^2}$, $\boldsymbol{D}$ is the first difference matrix such that $(\boldsymbol{D}\vec{x})_u = \vec{x}_u - \vec{x}_{u-1}$, the $\circ$ operator is the Hadamard product of two vectors (i.e. the element-wise product), and $\gamma_1, \gamma_2$ are the regularization parameters, discussed more below. The minimization is implemented numerically using the CVXPY library \cite{diamond2016cvxpy}.

The terms in Equation \ref{eq:bandpass_minimizer} are motivated as follows:
\begin{enumerate}
    \item Data Fidelity: The first term ensures the retrieved solution remains consistent with the observed data.
    \item Weighted Tikhonov Regularizer: The second term penalizes deviations from the prior in regions where the responsivity is expected to be zero.
    \item Weighted Shifted Laplacian Regularization: The third term enforces smoothness on the difference between the prior and the result. However, the enforcement strength is weighted by the prior, causing the regularizer to prioritize smoothness in high-responsivity regions. In areas where the prior approaches zero, the solution is allowed greater flexibility to maximize data fidelity, with the expectation that any resulting artifacts can be addressed via post-process smoothing.
\end{enumerate}

In practice, the regularization parameters $\gamma_1$ and $\gamma_2$ act as tuning dials that control a critical trade-off: how closely the algorithm should follow noisy measurements versus how strongly it should enforce our baseline expectations (i.e., how closely the retrieved responsivity should follow the prior). To determine the optimal values, we performed a systematic empirical search, testing various combinations to identify the pairing that consistently yielded the lowest percent error across all trials. Rigorous tuning ensures the algorithm generalizes reliably across all intended use cases rather than overfitting to a specific set of observations. Because the structure of the prior $\vec{p}$ heavily influences the ideal parameter values, we tuned the regularizers independently for each filter. Ultimately, we found that combining both regularizers yields the highest quality results; validation indicates that placing a significantly higher weight on $\gamma_2$ relative to $\gamma_1$ produces the most stable solutions.

\begin{figure}[hbtp]
    \centering
    \includegraphics[width=0.98\linewidth]{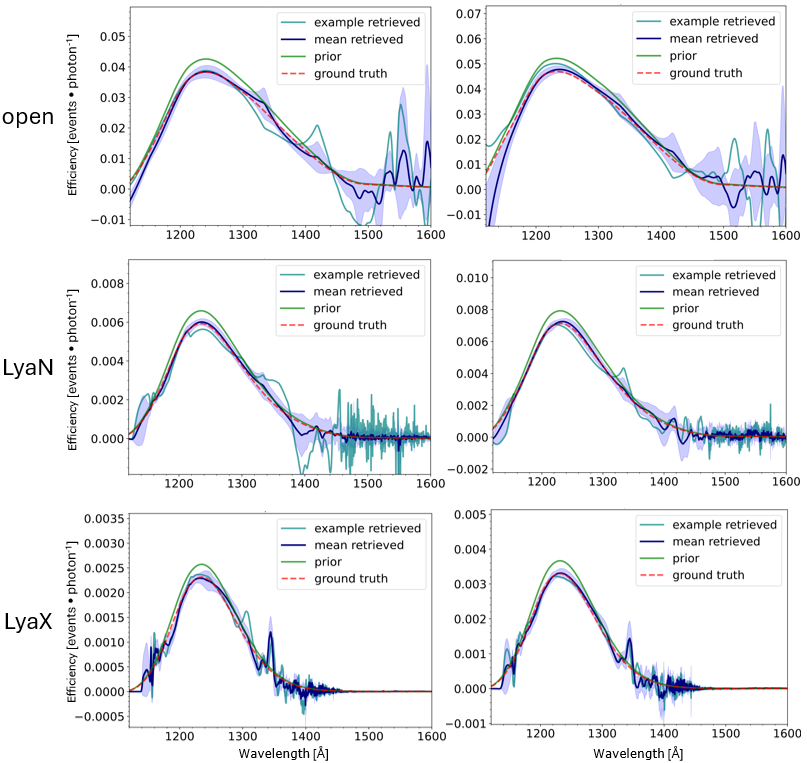}
    \caption{Validation results of the responsivity retrieval algorithm for both channels (columns) and open/LyaN/LyaX filter configurations (rows) across 50 independent trials. Filter efficiency $\varepsilon(\lambda)\tau_b(\lambda)$ is plotted instead of responsivity $r_b(\lambda)$ in order to match Figure \ref{fig:lab_responsivity}; the two are related via known scalar multipliers given in Equation \ref{eq:resp_funct_def}. In each subplot, the retrieved mean efficiency (dark blue solid line) closely tracks the ground truth (red dashed line), demonstrating the algorithm's ability to retrieve efficiency accurately despite a prior that does not match the ground truth (green solid line). A single example retrieval (light blue solid line) illustrates the typical trial-to-trial fluctuation, which becomes highly variable at longer wavelengths. The shaded blue region represents the uncertainty surrounding the mean retrieved value. \label{fig:retrieval_bandpass}}
\end{figure}

Validation results are illustrated in Figure \ref{fig:retrieval_bandpass}. Overall, the retrievals accurately track the ground truth until the responsivity nears zero. In these low-signal regions, the solution becomes noisy as the minimizer attempts to satisfy data fidelity while constrained by the regularizers. These high-frequency fluctuations can be corrected with a final smoothing pass and are thus acceptable.  For quantitative validation, we rely on the science requirements traceability matrix, which specifies a required Ly-$\alpha$ responsivity retrieval accuracy of $25\%$ and precision of $10\%$. In terms of the validation framework described here, these requirements imply that the mean retrieval error must not exceed $25\%$, and the standard deviation across retrievals must remain within $10\%$. This criterion applies across all bandpass filters to guarantee the mission's fundamental scientific objectives.  While there are no formal accuracy or precision requirements on the responsivity retrieval at 1304{\AA} and 1356{\AA}, since these wavelengths primarily inform the out-of-band photon background removal algorithms detailed in Zhang et al. \cite{Zhang26a}, we adopt the Ly-$\alpha$ retrieval requirement as a consistent, conservative benchmark to evaluate algorithmic stability at these wavelengths. Because the responsivities at 1304{\AA} and 1356{\AA} in the WFI filters and in the NFI LyaX filter are not utilized for mission science, their statistics are omitted from this analysis.

Table \ref{tab:bandpass_perc_err} summarizes the mean and standard deviation of the percentage errors for the retrieved responsivities at the channel/filter/wavelength combinations of scientific interest. These results reflect 50 independent trials for each bandpass filter, following the validation framework established in Section \ref{sec:validation_framework}. As shown, the retrieval errors for all evaluated bandpass filters comfortably satisfy the required performance thresholds.

\begin{table}[htpb]
    \centering
    \caption{Retrieved responsivity percent error statistics over 50 trials for bandpass filters. In the table heading, $\hat{\mathbb{E}}[\vec{E}_{\lambda}]$ denotes the empirical mean percent error of the retrieved responsivity at wavelength $\lambda$, while $\hat{\sigma}[\vec{E}_{\lambda}]$ denotes the empirical standard deviation of the retrieved responsivity at wavelength $\lambda$. See Equation \ref{eq:perc_err_def} for the definition of $\vec{E}$. The 'REQ' row denotes the maximum allowable error thresholds for Ly-$\alpha$ (1216{\AA}) retrievals as dictated by mission requirements. Channel/filter/wavelength combinations not utilized for mission science are excluded.}
    \label{tab:bandpass_perc_err}
    \begin{tabular}{ll|rrrrrr}
        \hline
        \textbf{Channel} & \textbf{Filter} & $\hat{\mathbb{E}}[\vec{E}_{1216}]$ & $\hat{\sigma}[\vec{E}_{1216}]$ & $\hat{\mathbb{E}}[\vec{E}_{1304}]$ & $\hat{\sigma}[\vec{E}_{1304}]$ & $\hat{\mathbb{E}}[\vec{E}_{1356}]$ & $\hat{\sigma}[\vec{E}_{1356}]$ \\
        \hline
        REQ & REQ & 25 & 10 & - & - & - & - \\
        \hline
        NFI & open & 0.05 & 4.14 & -2.97 & 3.93 & -2.65 & 11.06 \\
        NFI & LyaN & 0.11 & 3.81 & 0.19 & 5.62 & -1.96 & 16.80 \\
        NFI & LyaX & 0.86 & 3.93 & - & - & - & - \\
        WFI & open & 0.02 & 3.99 & - & - & - & - \\
        WFI & LyaN & -0.09 & 4.15 & - & - & - & - \\
        WFI & LyaX & -0.33 & 4.43 & - & - & - & - \\
        \hline
    \end{tabular}
\end{table}

\subsection{Algorithms and Results: Longpass Filters}
\label{sec:results_longpass}

For the longpass filters, the expression to be minimized is given as
\begin{equation}
\label{sec:longpass_equation}
\hat{\vec{r}} = \argmin_{\vec{r}} |\vec{S} - \boldsymbol{F}\vec{r}|_2^2 + \gamma |\vec{p} \circ \boldsymbol{D}(\vec{r}- \vec{p})|_2^2
\end{equation}

\begin{figure}
    \centering
    \includegraphics[width=0.99\linewidth]{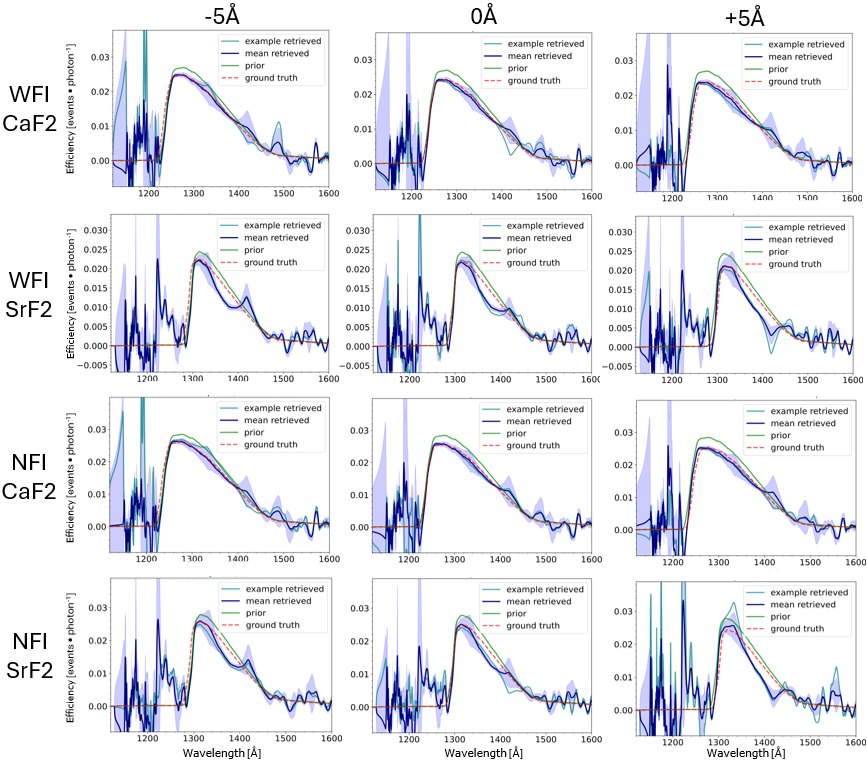}
    \caption{Validation results of the responsivity retrieval algorithm for both channels on the CaF2/SrF2 filter configurations (rows) and across $0${\AA} and $\pm5${\AA} ground-truth shifts (columns) for 50 independent trials, which simulate the expected thermal drift of the crystalline fluoride filter edges. Filter efficiency $\epsilon(\lambda)\tau_b(\lambda)$ is plotted instead of responsivity $r_b(\lambda)$ in order to match Figure 1; the two are related via known scalar multipliers given in Equation 2. In each subplot, the retrieved mean efficiency (dark blue solid line) closely tracks the shifted ground truth (red dashed line), demonstrating the algorithm's ability to retrieve efficiency accurately despite a stationary prior that does not match the ground truth (green solid line). A single example retrieval (light blue solid line) illustrates the typical trial-to-trial fluctuation, which becomes highly variable outside the primary transmission passband. The shaded blue region represents the uncertainty surrounding the mean retrieved value.}
    \label{fig:retrieval_longpass}
\end{figure}

\noindent As with the bandpass retrieval cost function in Equation 
\ref{eq:bandpass_minimizer}, $\boldsymbol{D}$ is the first difference matrix such that $(\boldsymbol{D}\vec{x})_u = \vec{x}_u - \vec{x}_{u-1}$, and $\gamma$ is the regularization parameter. The regularization approach also employs the same weighted shifted Laplacian described in the previous section. However, for the longpass filters, validation testing indicated that the inclusion of a Tikhonov regularizer negatively impacted the accuracy of the recovered solutions. Consequently, that term is omitted for the retrieval of optical responsivity for the longpass filters. Performance results for this algorithm are presented in Figure \ref{fig:retrieval_longpass}.

Consistent with previous findings, retrieval performance remains robust until the responsivity approaches zero, where numerical instabilities introduce heightened variance into the solutions. Furthermore, the $\text{SrF}_2$ filter configurations across both channels exhibit a distinct spectral artifact (a localized dip at approximately $1380$\AA), which is likely attributable to the stellar target selection process. While this feature presents an opportunity for subsequent algorithmic refinement, it does not compromise the scientific objectives of the Carruthers mission. Similarly, although the algorithm predicts the temperature-induced spectral edge shifts, which stem from thermal variations within the crystalline fluoride substrates, with limited precision, this minor discrepancy remains well within acceptable tolerances for the mission's overarching requirements.

Regarding quantitative validation, longpass filters possess negligible responsivity at $\text{Lyman-}\alpha$ \textit{a priori}; consequently, standard $\text{Ly-}\alpha$ accuracy requirements are inapplicable. Although formal accuracy thresholds are absent for the $1304${\AA} and $1356${\AA} wavebands of the NFI $\text{CaF}_2$ and $\text{SrF}_2$ configurations, given that these measurements primarily support the background-removal algorithms detailed in Zhang et al. \cite{Zhang26a}, we adopt the rigorous $\text{Lyman-}\alpha$ requirement used to assess the bandpass filter performance as a conservative metric for evaluating algorithmic stability.

Table \ref{tab:longpass_perc_err} summarizes the mean and standard deviation of the percentage errors for the retrieved responsivities at the channel/filter/wavelength combinations of scientific interest.  While the corresponding performance curves for the WFI longpass channels are still displayed graphically, their quantitative error statistics are excluded from the evaluation tables, as these specific bandpass combinations are not utilized for primary mission science objectives. These results reflect 50 independent trials for each longpass filter, following the validation framework established in Section \ref{sec:validation_framework}. As shown, the retrieval errors for all evaluated longpass filters comfortably satisfy the required performance thresholds.

\begin{table}[htpb]
    \centering
    \caption{Retrieved responsivity percent error statistics over 50 trials for longpass filters. In the table heading, $\hat{\mathbb{E}}[\vec{E}_{\lambda}]$ denotes the empirical mean percent error of the retrieved responsivity at wavelength $\lambda$, while $\hat{\sigma}[\vec{E}_{\lambda}]$ denotes the empirical standard deviation of the retrieved responsivity at wavelength $\lambda$. See Equation \ref{eq:perc_err_def} for the definition of $\vec{E}$. Channel/filter/wavelength combinations not utilized for mission science are excluded.}
    \label{tab:longpass_perc_err}
    \begin{tabular}{ll|rrrr}
        \hline
        \textbf{Channel} & \textbf{Filter} & $\hat{\mathbb{E}}[\vec{E}_{1304}]$ & $\hat{\sigma}[\vec{E}_{1304}]$ & $\hat{\mathbb{E}}[\vec{E}_{1356}]$ & $\hat{\sigma}[\vec{E}_{1356}]$ \\
        \hline
        NFI & CaF2 & 0.45 & 6.11 & 4.41 & 7.09 \\
        NFI & SrF2 & 1.98 & 5.06 & 14.32 & 6.71 \\
        \hline
    \end{tabular}
\end{table}

\section{Conclusion}

The Carruthers Geocorona Observatory successfully launched in September 2025 and commenced science operations at the Earth-Sun L1 Lagrange point in March 2026. A critical prerequisite for deriving scientific insights from raw instrument data is the precise characterization of optical system responsivity, and we present a novel approach to on-orbit stellar calibration that is capable of highly accurate estimation of responsivity as a function of wavelength across the sensor passband.  Validation efforts using synthetic stellar observations demonstrate high fidelity at the Lyman-$\alpha$ line in particular. Driven by the mission's autonomous pointing control and robust regularization techniques in the inverse retrieval process, the percent error statistics for the LyaN science filters are $0.11\% \pm 3.81\%$ for the Narrow Field Imager (NFI) and $-0.09\% \pm 4.15\%$ for the Wide Field Imager (WFI). Because these evaluations fall well below the baseline requirement threshold of $25\% \pm 10\%$ calibrated radiance accuracy, the proposed science data processing pipeline successfully guarantees high-confidence measurements of the geocorona.

\subsection*{Acknowledgements}
This project is based on observations made with IUE, obtained from the Mikulski Archive for Space Telescopes.

This project has made use of the SIMBAD database, operated at Centre de Données astronomiques de Strasbourg (CDS), France.

\begin{itemize}
\item Funding: This work was supported by the NASA Science Mission Directorate, Heliophysics Division through contract 80GSFC21C0038.
\item Conflict of interest/Competing interests:
Not applicable
\item Ethics approval and consent to participate:
Not applicable
\item Consent for publication:
Not applicable
\item Author contribution: Ordered in author list.
\end{itemize}


\bibliographystyle{ieeetr} 
\bibliography{ref}

\end{document}